\documentclass[10pt, conference]{IEEEtran}
\ifCLASSINFOpdf
  \usepackage[pdftex]{graphicx}
\else
\fi
\begin{document}
\title{Piggybacking on an Autonomous Hauler: Business Models Enabling a System-of-Systems Approach to Mapping an Underground Mine}


\author{
\IEEEauthorblockN{Markus Borg}
\IEEEauthorblockA{Software and Systems Engineering Lab.\\
RISE SICS AB\\
Lund, Sweden\\
markus.borg@ri.se}
\and
\IEEEauthorblockN{Thomas Olsson}
\IEEEauthorblockA{Software and Systems Engineering Lab.\\
RISE SICS AB\\
Lund, Sweden\\
thomas.olsson@ri.se}
\and
\IEEEauthorblockN{John Svensson}
\IEEEauthorblockA{Link\"{o}ping University\\
Boliden Mineral AB\\
Boliden, Sweden\\
john.svensson@boliden.com}
}


%Telephone: (800) 555--1212, Fax: (888) 555--1212}


\maketitle

\begin{abstract}
% HIGHLY CONDENSED VERSION OF THE RESEARCH. Can also cycle I-M-R-M-R-D or I-M-R-D-R-D.
%
% Introduction move
%	- Background step
%	- Objectives step
%	- Results step
%	- Scope step
%
% Methods move
%	- Merely mention method step
%	- Brief description of the method step
%
% Results move
%	- Present new data step
%	- Present novel method step
%	- Present new model step
%	- Present validation of the model step
%
% Discussion move
%	- Principal conclusions about the outcomes step
%	- Implications of the outcomes step
%	- Applications step
%	- Interpretations step		
With ever-increasing productivity targets in mining operations, there is a growing interest in mining automation. 
In future mines, remote-controlled and autonomous haulers will operate underground guided by LiDAR sensors.
We envision reusing LiDAR measurements to maintain accurate mine maps that would contribute to both safety and productivity. 
Extrapolating from a pilot project on reliable wireless communication in Boliden's Kankberg mine, we propose establishing a system-of-systems (SoS) with LIDAR-equipped haulers and existing mapping solutions as constituent systems.
%As SoS relies on reliable wireless communication, we base our proposal on pilot 
SoS requirements engineering inevitably adds a political layer, as independent actors are stakeholders both on the system and SoS levels.
We present four SoS scenarios representing different business models, discussing how development and operations could be distributed among Boliden and external stakeholders, e.g., the vehicle suppliers, the hauling company, and the developers of the mapping software. 
Based on eight key variation points, we compare the four scenarios from both technical and business perspectives.
Finally, we validate our findings in a seminar with participants from the relevant stakeholders.
We conclude that to determine which scenario is the most promising for Boliden, trade-offs regarding control, costs, risks, and innovation must be carefully evaluated.

%The PIMM project addresses the fundamental challenge of network communication by constructing a pilot 5G network in the underground mine Kankberg. 
%In this report, we discuss how such a 5G network could constitute the essential infrastructure needed to organize existing systems in Kankberg into a system-of-systems (SoS). 
%Our proposal is to connect LiDAR-equipped vehicles operating in the mine with existing mine mapping solutions. 
%The approach is motivated by the approaching era of remote controlled, or even autonomous, vehicles in mining operations. 
%The proposed SoS could ensure continuously updated maps of Kankberg, rendered in unprecedented detail, supporting both safety and productivity in the underground mine. 

\end{abstract}

\begin{IEEEkeywords}
system-of-systems, business models, architecturally significant requirements, software ecosystems, mining automation.
\end{IEEEkeywords}

% \fi
\IEEEpeerreviewmaketitle

\section{Introduction}
% PROVIDE AN ACCOUNT OF THE NEW RESEARCH. MAKE CLEAR IT IS NOVEL AND SIGNIFICANT.
%
% Establish a territory move (link the problem studied to the general research area)
%	- Claim centrality step (this is important to us all because...)
%	- Indicate a territorial lack or problem step
%	- Present background/topic generalizations step
%	- Review previous research step
%
% Establish a niche move (To indicate there is a need within the research area)
%	- Indicate lacks in previous research step
%	- Indicate that the present research follows a tradition step
%	- Indicate an unsolved research or real-world problem
%	- Make counterclaims step (very strong to do)
%
% Occupy the niche move
%	- Announce objectives step
%	- Announce principal findings step
%	- Announce or briefly describe the scope of the present research step
%	- Refer to important aspects step
%	- Announce the RA structure step

The global demand for raw materials is increasing, thus mining operations are producing close to their capacity limits. 
Continuous operation is essential to successful mining, any production losses involve considerable monetary consequences. 
With ever-increasing productivity targets, there is a growing interest in mining automation.
Safety, however, is still the primary concern of any mining operation, i.e., improving worker safety trumps reaching increased productivity goals. 
In line with expectations on digitalization in industry in general, mining operations hope to experience a considerable paradigm shift through increased connectivity in the mines.
Reliable wireless communication would enable both advanced data analytics and the advent of remote controlled, or even autonomous, heavy equipment such as LHD vehicles (load, haul, dump).
 
As a byproduct of future autonomous vehicles operating in underground mines, there will be vehicles continuously using LiDAR measurements of the mine. 
In this report, we discuss the potential to use these LiDAR data to maintain a detailed 3D map, i.e., enabling online availability of continuously updated geospatial information of drifts and crosscuts. 
We propose combining LiDAR-equipped LHD loaders with Boliden's current mapping and positioning solutions into a system-of-systems (SoS). 
SoS is defined by Kotov as ``large-scale concurrent and distributed systems of which the components are complex systems themselves''~\cite{kotov_systems_1997}. 
In a SoS, the constituent systems accomplish their own goals (i.e, by composing their respective elements into a system~\cite{axelsson_systems--systems_2015}), but the SoS functionality is more than the sum of its constituent systems. 
A key characteristic of a SoS is the independence of its constituent systems, i.e., they are typically developed independently, but through evolutionary development they are organized in a SoS, allowing novel goals to be met through emergent behavior~\cite{maier_architecting_1996}.

Our proposal assumes access to reliable wireless communication as studied in the PIMM project, Pilot for Industrial Mobile communication in Mining\footnote{https://www.sics.se/projects/pimm}, exploring deployment of a 5G communication network in Boliden's Kankberg underground mine.
This paper is based on a technical report describing the case in detail~\cite{borg_lidar_2017}.

This paper is guided by two research questions (RQ): 
\begin{itemize}
\item RQ1. What business models could enable a mapping SoS in Kankberg?
\item RQ2. From Boliden's perspective, how do the SoS requirements vary depending on the business model? 
\end{itemize}
Based on the RQs, we discuss four alternative business models: 1) an in-house solution, 2) buying a mapping service as an add-on feature from the vehicle suppliers, 3) buying a mapping service as an additional service from the hauling service provider, and 4) a solution based on software ecosystem~\cite{jansen_software_2013}.
For each scenario, we present a high-level SoS architecture and discuss eight key variation points covering both technical and business perspectives. 
%The amount of effort required by Boliden is discussed, separating development and operations. 

The rest of the paper is organized as follows: Section~\ref{sec:bg} presents the background and context, as well as related work on mine mapping.
Section~\ref{sec:techbg} introduces the fundamentals of the proposed SoS, and describes the assumptions we make for all four scenarios. 
Sections~\ref{sec:scen1}-\ref{sec:scen4} contain the main contribution: a presentation of the four SoS scenarios, and Section~\ref{sec:disc} discusses our findings.
Finally, Section~\ref{sec:sum} contains a summary, and recommendations for future work toward realizing a SoS.

\section{Background and Related Work} \label{sec:bg}
This section covers Boliden background information and a description of the pilot mine Kankberg. 
We also introduce mining safety and some ongoing work on remote controlled and autonomous LHDs.
Finally, we present state-of-the-art mine mapping using robots from the research front.

\subsection{Boliden and the Kankberg Mine}
Boliden is a leading metals company with core competency within exploration, mining, smelting, and metal recycling. 
The company has approximately 5,500 employees. 
Boliden's success in the metals business relies on state-of-the-art technology, including efficient mine design, mobile control systems, and increased automation; the latter being a strong trend in mining operations globally~\cite{bellamy_assessing_2011,gustafson_influence_2013}. 

Boliden has an explicit ambition to run projects in-house, i.e., internal know-how is considered fundamental. 
The current automation trend promises increased productivity through the introduction of autonomous machines, wireless data transfer, and positioning of people and equipment –- all these solutions rely on software-intensive systems. 
However, whether Boliden will be able to keep all required technical software expertise in-house is uncertain. 
The scaling role of software in traditional industries has attracted considerable research efforts lately~\cite{fitzgerald_scaling_2016}, and turning into a software-intensive company is an acknowledged challenge. 
The IT department at Boliden roughly employs 100 people including support functions, but few of them are software developers.

\begin{figure}
\centering
\includegraphics[width=0.5\textwidth]{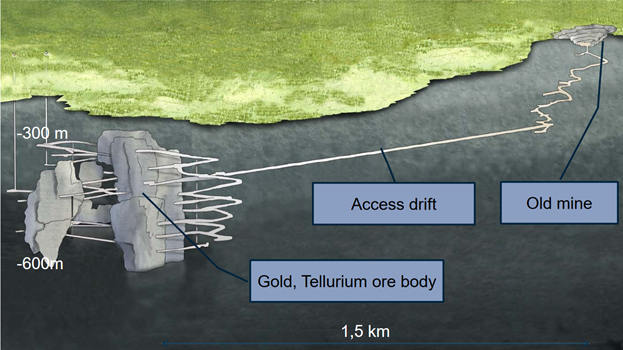}
\caption{Overview of the Kankberg mine.}
\label{fig:Kankberg}
\end{figure}

The Kankberg mine was closed in the 1990s, but it was reopened in 2012 due to the discovery of a new gold and tellurium ore body.
In the mine, the cut-and-fill method is used with granite as the backfill material. 
Drilling and blasting is used to break rock for excavation, i.e., controlled use of explosives. 
Kankberg employs about 80 people and 20 contractors.
Fig.~\ref{fig:Kankberg} shows an overview of the Kankberg mine. 
In mining terminology, \textit{shafts} are vertical tunnels used for ventilation and transportation. 
Horizontal tunnels are called \textit{drifts} (cf. the main access drift in Fig.~\ref{fig:Kankberg}). 
Drifts across the ore body are referred to as \textit{crosscuts}. 

\subsection{Mining Safety and the Importance of Mine Mapping}
Safety is the top priority in any mining operation. 
Boliden has a zero vision for accidents in 2018, and is determined to use the best available technology and processes to reach the goal. 
Several risks associated with working in an underground mine must be considered, including cave-ins, flooding, gas explosions, chemical leakage, and electrocution, and also long-term health risks related to mineral dust, radon, welding fumes, mercury, noise, and heavy loads. 
The SoS envisioned in this paper primarily focuses on threats related to cave-ins.

\begin{figure}
\centering
\includegraphics[width=0.5\textwidth]{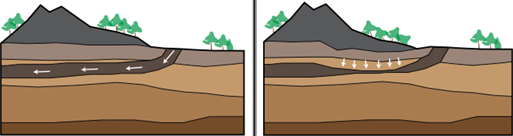}
\caption{Subsidence caused by underground mining. Illustration from Wikimedia Commons by Mpetty1 (Own work, CC BY-SA 3.0).}
\label{fig:subsidence}
\end{figure}

As the mining operations proceed, shafts and drifts are inevitably affected in position (inclination, rotation, lateral movements, and curves) and form (compression and deformation). 
There are three main causes of cave-ins in underground mines. 
First, hasty mining operations might fail to secure walls and ceilings of shafts and drifts. 
Second, excessive excavation might lead to cracks in floors and walls, thus weakening the entire structure. 
Examples include insufficient vertical spacing between crosscuts and too rectangular crosscuts causing stress concentration in corners. 
Third, gradual sinking of land can cause cave-ins~\cite{thrun_autonomous_2004}, i.e., subsidence (the downward motion of a surface). 
As illustrated in Fig.~\ref{fig:subsidence}, mining operations induce subsidence of the Earth's surface. 
While mining-induced subsidence is rather predictable in magnitude and extent, monitoring the progress is fundamental to mining safety. 
In Kankberg, however, the mountain stresses caused by drilling and blasting activities dominate any subsidence.  

Currently, Boliden's primary equipment to perform stress-strain measurements in Kankberg are extensometers and cable bolts. 
In areas directly impacted by drilling and blasting, optical measurements are also conducted, i.e., manually collected data comparing distances to reference points with a millimeter precision. 
Regular shaft inspections are mandated by mine safety regulations, i.e., the determination of spatial changes of shaft columns that could be indicative of cave-ins. 
However, shaft inspections traditionally are time-consuming, and require halting of the mining operations.
Furthermore, while the shaft inspections are critical, there is also a need to survey drifts and crosscuts.
Laser scanning of mine shafts is already an area for which several companies offer services, e.g., SightPower\footnote{http://sight-power.com/en/solutions/mineshaft-inspection/} and DMT\footnote{http://www.dmt-group.com/en/services/exploration/surveying-geoinformation.html}, but measuring horizontal structures has received considerably less attention. 

One promising approach to monitor changes in rock mass plasticity, originating in either drilling and blasting or subsidence, is to maintain highly accurate maps of the underground mine. 
However, for the maps to be truly useful from a safety perspective, such maps need to be as precise as current optical measurements. 
We believe that superimposition of subsequently scanned 3D images, i.e., LiDAR point clouds, could be used to determine and highlight issues such as ground movements, cracks in lining, and misalignment, analogous to what is currently the state-of-the-art offer for shaft inspections. 
%Another reason to maintain accurate maps is commercial. 
%The creation of volumetric maps is critical to the mining business, i.e., maps that estimate the value of an ore body.

Future autonomous LHDs operating underground will rely on LiDAR for navigation, a technology already used in numerous applications, from construction industry and road maintenance to city management and planning~\cite{schwartz_lidar:_2010}.
LiDAR allows high accuracy measurements of the coordinates of the point in space where a laser beam reflects from a non-transparent object. 
While mine environments present certain challenges for laser scanning systems~\cite{huber_automatic_2006}, e.g., widely-ranging albedo, shiny metal, and wet surfaces, for the purposes of our SoS discussion, we posit that future LiDAR systems will meet the accuracy requirements for a mapping solution.

% candidate to remove I propose.... 
\subsection{Robotic Mine Mapping}
Several researchers have worked on mine mapping using autonomous robots. 
The research has mainly targeted abandoned mines, a substantial problem in the US; not even the US Bureau of Mines are certain of how many abandoned mines exist in the country. 
Accurate maps of the old mines are typically missing, but could be of great value to avoid catastrophic events~\cite{pauley_preliminary_2002}. 
The motivation for fully autonomous robots has been twofold. 
First, abandoned mines present harsh environments, dangerous to humans. 
Dangers include lack of structural soundness, low oxygen levels, and risk of flooding. 
Second, the inadequacy of current wireless communication techniques makes remote controlled robots unfeasible.

Thrun and colleagues have conducted several experiments on autonomous mapping of abandoned mines. 
Their robot Groundhog evolved into a 1,500-pound robotic vehicle, equipped with, e.g., onboard computing, laser range sensing, gas sensor, and a video recorder~\cite{thrun_autonomous_2004}. 
They have reported a series of successful mapping operations~\cite{baker_campaign_2004}. 
However, limited by the technology at the time, the robot did only perform 3D scans at regular vantage points; using a tilting mechanism to acquire a point cloud of the area in front of the robot. 
Current autonomous vehicles acquire 3D point clouds more frequently.

Huber and Vandapel and their research group also did work on underground mine mapping using robots~\cite{huber_automatic_2006}. 
In contrast to the work by Thrun \textit{et al.}, Huber and Vandapel tried their approach in active mines. 
They mounted a high-resolution 3D scanner on a mobile robot, providing 8000 x 1400 pixel scans with millimeter-level accuracy. 
As for Groundhog, they collected scans only at certain vantage points; 
Each three to five meters the robot stopped for 90 seconds to obtain a complete scan. 
A considerable contribution of their research is related to multi-view surface matching, i.e., merging multiple 3D views into a single map. 
Their approach is called iterative merging, which was successfully used to create high-quality maps of an underground mine. 
However, their approach does not scale to large numbers of scans; 
Back in 2006, their approach could only generate sub-maps containing about 50 scans.

\section{Preconditions and Analysis Approach} \label{sec:techbg}
%\section{Technical Background: Introduction to a Potential Solution} \label{sec:techbg}
In this section, we introduce the preconditions and general issues that the envisioned SoS must address. 
The main challenge in designing and realizing the SoS is to organize the responsibilities of Boliden and their partners, incl. software vendors, vehicle suppliers, and service providers. 
In this paper, we distinguish between development (Dev) and operations (Ops)~\cite{bass_devops:_2015} -- and discuss pros and cons of different constellations and their corresponding business models.

\begin{figure}[t]
\centering
\includegraphics[width=0.5\textwidth]{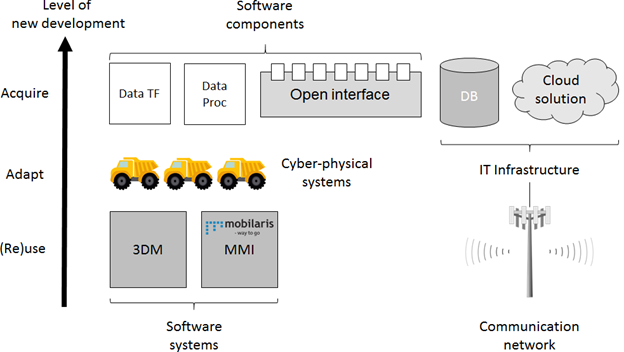}
\caption{Overview of the SoS parts involved in the solution proposals. Constituent software systems are presented as dark gray boxes while software components are white.}
\label{fig:infrastructure}
\end{figure}

Fig.~\ref{fig:infrastructure} lists all assets that are involved in the SoS, organized into the three dimensions acquire, adapt, and (re)use~\cite{papatheocharous_decision_2015}. 
The assets are of the following types: software systems, software components, cyber-physical systems, and IT infrastructure. 
Assets are organized along an y-axis depicting the level of new development required. 
Our solution proposals assume a reliable high-bandwidth wireless communication network in the Kankberg mine, e.g., a 5G public mobile network as piloted in the PIMM project, and that LiDAR equipped LHDs are operating in the mine.
Furthermore, our proposals rely on two existing systems already in use: 1) 3D Map rendering software (3DM), a commercial off-the-shelf 3D CAD solution used by the mining engineers, and 2) Positioning System (Pos), a positioning system for vehicles, equipment and personnel. 
Already today, Pos monthly imports map updates from 3DM.
%Also, existing LHDs that need to be adapted are presented, as well as IT infrastructure that either is reused or acquired.

%For all four scenarios, we make the following assumptions:
%\begin{itemize}
%\item Boliden owns or leases the vehicles but not necessarily all services around them.
%\item Reliable high-bandwidth 5G data communication is available in the Kankberg mine.
%\item LiDAR equipped LHDs from different vehicle suppliers operate in the underground mine.
%\item The format of the LiDAR data, i.e., detailed 3D point clouds, is not standardized, but can be transformed into a valid 3DM import format.
%\end{itemize}

Several alternative approaches to realize a SoS in Kankberg are possible. 
We present four contrasting scenarios originating in discussions with Boliden: 1) an in-house solution, 2) buying mapping capability as an add-on feature from the vehicle suppliers, 3) paying for mapping as an additional service from the hauling service provider, and 4) a software ecosystem.
We analyzed the four scenarios by exploring a future Kankberg SoS from the perspective of seven central SoS questions, stated by Axelsson in the strategic SoS research and innovation agenda for Sweden~\cite{axelsson_systems--systems_2015}.
We found that the answers to five of the seven questions would be the same across the scenarios:

\textbf{1) \textit{Why} is it created?} The new SoS is created to improve safety and productivity in the underground mine. 
The SoS should leverage on LiDAR equipped LHDs operating in the mine, and reuse the obtained 3D point clouds to update 3DM and in turn Pos. 

\textbf{2) \textit{Whose} system is it?} Boliden takes ownership of the SoS. 
%Agreements must be made with the vehicle suppliers to acquire 3D point clouds from the LiDAR equipped LHDs.

\textbf{3) \textit{Who} are the stakeholders?} Boliden is the primary beneficiary of the new SoS. 
Secondary beneficiaries include stakeholders of the constituent systems: vehicle suppliers could improve their predictive maintenance, and the enterprise offering Pos will benefit from pinpointing of equipment and personnel on exact maps. 
%Furthermore, safer working conditions are in the interest of the labor unions.

\textbf{4) \textit{What} should it do?} The SoS should enable continuous updates of Kankberg maps at unprecedented level of detail. 

\textbf{5) \textit{How much} should it perform?} The main quality attributes of the SoS is performance and mapping accuracy. 
The mapping accuracy should be on a sub-centimeter scale, and the computational performance should allow map updates at least on an hourly rate.

Regarding the two remaining central SoS questions, we found that the scenarios differ considerably.
\textbf{6) \textit{How} should it be organized?} and \textbf{7) \textit{When} does it change?}
Based on these two questions, we analyzed how the four scenarios differed.
We elicited variation points, and iteratively condensed them into eight \textit{key variation points} -- three of them technical, the rest geared towards business perspectives, see Table~\ref{tab:keyvp}.
Consequently, in Sections~\ref{sec:scen1}-\ref{sec:scen4}, we discuss the four scenarios based on the following eight key variation points: IT infrastructure, Evolution, Architecturally significant requirements; Inter-company dependency, Upfront investment, Running costs, Risks, and Innovation Platform.

\begin{table}
\centering
\caption{Key variation points discussed for each scenario.}
\label{tab:keyvp}
\begin{tabular}{|p{0.2\columnwidth}|p{0.7\columnwidth}|}
\hline
\multicolumn{2}{|c|}{Technical perspective} \\
\hline \hline
IT infrastructure & Databases, servers, network, etc. Also cloud components, network requirements, storage, etc.  \\
\hline
Evolution & Decision whether to make (develop) internally or buy, scalability and control of roadmap.\\
\hline
ASRs                     & Architecturally Significant Requirements (ASRs) for Boliden, namely security, reliability, and efficiency. \\
\hline\hline
\multicolumn{2}{|c|}{Business perspective} \\
\hline\hline
Inter-company dependency & Business relationships, such as whether there are dedicated solutions or generic solutions being used. \\
\hline
Upfront investment       & Fixed initial costs that are involved in developing the SoS. \\
\hline
Running costs            & Costs subscription fees, service costs or operations costs. \\
\hline
Risk                     & Risks associated with unsatisfactory operations of the SoS, i.e., an inadequate mapping solution.  \\
\hline
Innovation platform      & Strength of the innovation culture and processes from the SoS partners.\\
\hline
\end{tabular}
\end{table}

Regarding architecturally significant requirements (ASR), we provide a clarification of our intentions.
ASRs are the requirements that are more influential on the system architecture than others. 
From Clements and Bass~\cite{clements_relating_2010}:
\textit{``Most of what is in a requirements specification does not determine or -- `shape' an architecture. 
Architectures are mostly driven or shaped by quality attribute requirements. 
These determine and constrain the most important architectural decisions.''} 
%And yet the vast bulk of most requirements specifications is focused on the required features and functionality of a system, which shape the architecture the least. 
%Worse, most do a poor job of specifying quality attributes; many ignore them altogether.''}
Besides having a larger impact on the architecture, ASRs tend to be vague and context dependent~\cite{chen_characterizing_2013}. 
As pointed out by both Clements and Bass~\cite{clements_relating_2010} and Anish \textit{et al.}~\cite{anish_what_2015}, ASRs are often non-functional requirements (a.k.a. NFRs or quality requirements). In this paper, we focus our discussion on ASRs on the following three NFRs: 1) \textit{Security}, including integrity of data and users, access control (including identification and access logs, etc.) and transmission of data in a secure manner, 2) \textit{Reliability}, encompassing availability of a service, loss of data and time to recover from a failure are included, and 3) \textit{Efficiency}, e.g., response-time, use of resources (such as storage or battery), and bandwidth. 

Finally, we validated the feasibility of our findings by conducting a static validation, as recommended by Gorschek \textit{et al.}~\cite{gorschek_model_2006}.
We invited all members of the PIMM project to a teleconference seminar, i.e., Boliden, Volvo Construction Equipment, Ericsson, Telia, ABB, Lule\r{a} University of Technology, and Wolfit.
During the seminar, attended by roughly 15 participants, we introduced fundamental concepts of systems-of-systems and presented the four scenarios.
The seminar was recorded, and the feedback we received was used to finalize this paper and the more comprehensive technical report~\cite{borg_lidar_2017}.

%TO: Added subsection
%\subsection{The seven questions} \label{sec:7q}
%Each individual scenario can be realized in numerous ways, but for the clarity of this report, we present one complete possible design for each case. 
%On the other hand, certain aspects of the SoS design will be the same across all four scenarios. 
%We identify these by exploring the seven key questions stated by 
%We argue that the answers to the first five questions are the same:

%\subsection{Architecturally Significant Requirements}

\section{Scenario In-house} \label{sec:scen1}% - Boliden is System Integrator and Coordinator of IT Subcontractors}
In Scenario In-house, Boliden takes on the leading role in designing the SoS and acts as owner of the system integration.
While an IT subcontractor (Acme Consultants) does a large part of the actual development effort, Boliden undertakes initial requirements elicitation and contributes actively in the iterative specification of requirements.
Regarding the SoS evolution, however, the sub-contractor takes a leading role. 

\begin{figure}
\centering
\includegraphics[width=0.5\textwidth]{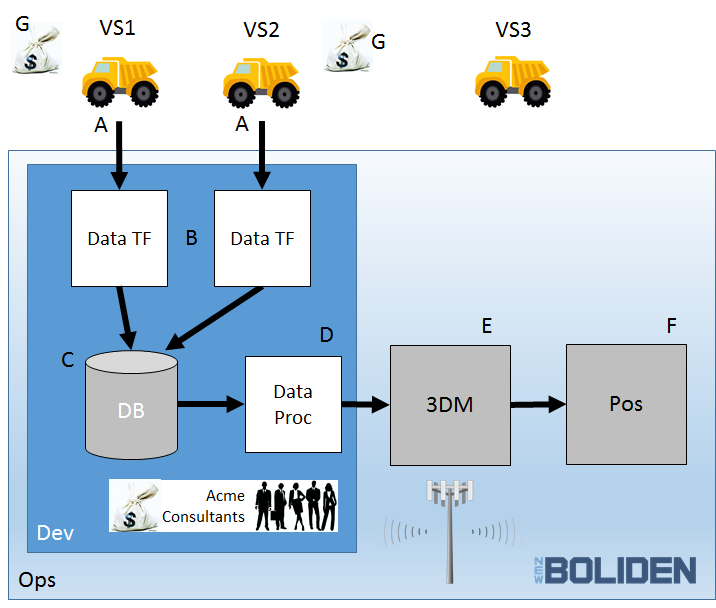}
\caption{Overview of Scenario In-house. Boliden owns both the system integration and the operations.}
\label{fig:scen1}
\end{figure}

Fig. \ref{fig:scen1} outlines the scenario. 
In the upper part of the figure, Boliden pulls the LiDAR data from the LHDs supplied by Vehicle Supplier 1 (VS1) and Vehicle Supplier 2 (VS2) through open APIs (cf. A). 
In this scenario, LHDs from Vehicle Supplier 3 (VS3) do not support access to any LiDAR data. Both LiDAR data from VS1 and VS2 are transformed, using two customized software components (cf. B), before being stored in the Boliden data storage (cf. C). 
The LiDAR data are then used in a data processing component (cf. D), filtering out measurements that differ from previous point clouds.
Moreover, the data are preprocessed into valid 3DM import format.
Finally, the LiDAR data are imported by 3DM to update its 3D CAD models (cf. E) – repeatedly synchronized with the maps in Pos (cf. F).

% Killing some darlings.... 
% Figure \ref{fig:scen1} also describes the scenario from a business perspective. Boliden acquires development resources from Acme Consultants, which means external development of the data transformations, the database solution, and the data processing (cf. area Dev) for an upfront investment. Boliden signs an agreement with VS1 and VS2 to allow extraction of the LiDAR data, realized through upfront investments (cf. G) but no running costs and no technical support. Apart from this agreement, the vehicle suppliers are not involved and do not obtain any direct benefits from the SoS. From an operations point of view, Boliden is solely responsible for the entire SoS, i.e., both the constituent systems and the new software components. Evolution of the SoS, including adapting to data and API changes of the LHDs, are not straightforward in this scenario as it requires additional resources acquired from Acme Consultants.

%Regarding the key variation points, we describe Scenario In-house as follows:
\subsection{Technical Key Variation Points}

\textbf{IT infrastructure.}
Boliden hosts all constituent systems using local resources. 
The systems are connected to the local high-bandwidth communication network and all data are kept on Boliden's local data storage. 
The LiDAR data are pulled from the vehicles by a system that also transforms the data into a common format. 
%Also computation is performed locally. 
%Most of the implementation is on systems under Boliden's control and will be tailored to the specific needs of Boliden. 

\textbf{Evolution.}
The resulting SoS will be proprietary to Boliden, although a subcontractor will perform most development.
As Boliden owns both the infrastructure and the SoS, the evolution pace is completely up to Boliden's willingness to invest. 
However, scalability might become an issue in this scenario, as there is a strong dependence on contractors, and scaling hardware means investing in more computers. 
Furthermore, as the solution is tailored to the specific needs of Boliden and the specific suppliers,  there is a risk that introducing or changing constituent systems will mean considerable development effort by Boliden. 
On a positive note, Boliden has complete control of the roadmap and the release schedule. 

\textbf{ASR.}
All computation and rendering of maps and usage of the data is performed by Boliden. 
No other company has access to any information, i.e., security is not a strong requirement in this scenario. 
In terms of scaling of the computation, Boliden is largely left on their own. 
Hence, efficiency requirements to ensure a scalable solution are important early on to avoid expensive refactoring late in the project or the need to add hardware later in the operations phase. % which might not even be possible without updating the software. 
Reliability is unlikely to be a driving requirement. 
While still important, the communication infrastructure is simpler in Scenario In-house compared to the others. 
%Furthermore, as Boliden operates all IT systems, their internal routines and procedures outside the actual software implementation should handle any availability problems. 
Consequently, efficiency is the most important ASR in this scenario. 

\subsection{Business Key Variation Points}

\textbf{Inter-company dependency.}
Boliden customizes data access to the LHDs supplied by VS1 and VS2. 
Moreover, the data processing is tailored to support the input format of 3DM. 
Consequently, the relationship deepens with selected companies, making Boliden more reliant on them and making it more difficult to switch both vehicle suppliers and mapping solution. 
The relationships tend to be static and stable. 

\textbf{Upfront investment.}
Boliden invests in internal solutions for data storage and computation as well as the development effort for the mapping SoS. 
Scenario In-house has the largest upfront investment.

\textbf{Running costs.}
Boliden staff or contractor need to ensure continued operation and basic maintenance. 
While there is a running cost, it is rather modest. 

\textbf{Risks.}
As responsible for operations, Boliden carries all risk and must ensure mitigation internally. 
Should there be a hardware problem, there is a risk of longer down-time. 
Boliden owns all collected LiDAR data, and the data are stored internally. 
While the LHD suppliers have access to the LiDAR data from their respective vehicles, the Scenario In-house still provides the highest data privacy among the four scenarios.

\textbf{Innovation platform.} 
There is little innovation from anyone else but Boliden. We assume that Acme is just supplying development resources, hence is unlikely to take an interest in driving innovation. As Boliden is in the mining industry and not software business, the innovation is customer-driven and not technology driven~\cite{bosch_speed_2016}. 

\section{Scenario Add-on Option} \label{sec:scen2}% - Vehicle Suppliers Sell the Mapping Feature as an Add-on Option}
In the second scenario, some vehicles suppliers have recognized mapping as a promising business opportunity.
The LiDAR sensors are mounted on their LHDs after all, why not make a profit by selling the LiDAR data to Boliden? 
VS1 and VS3 decide to offer the mapping feature as an LHD add-on option. 

\begin{figure}
\centering
\includegraphics[width=0.5\textwidth]{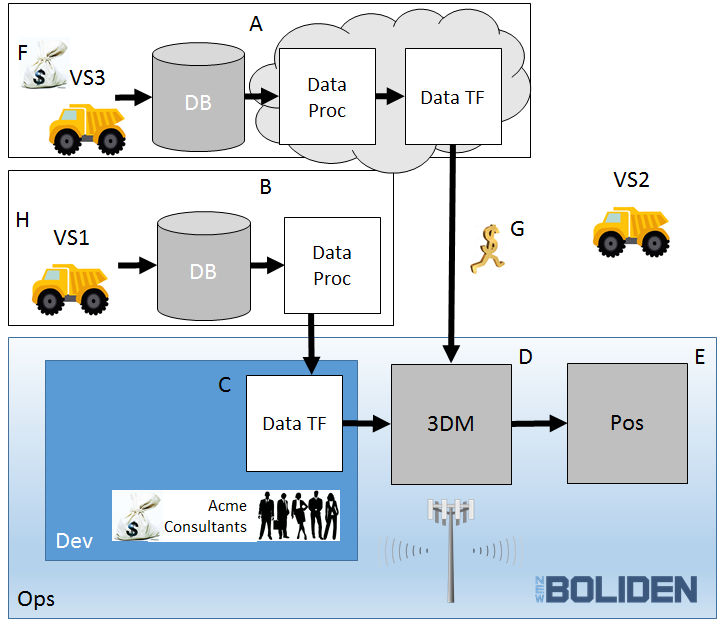}
\caption{Overview of Scenario Add-on Option. Some vehicle suppliers sell mapping as a feature, with or without data transformation for a subscription fee.}
\label{fig:scen2}
\end{figure}

Fig. \ref{fig:scen2} presents an overview picture of the scenario. 
The top of the figure shows VS3's LHDs storing LiDAR data locally. 
Each individual LHD from VS3 implements the entire pipeline (cf. A): data collection, data storage, data processing (incl. filtering), data transformation, and data transfer to Boliden's 3DM, i.e., both the data storage and computational resources are distributed. 
VS1 provides a similar add-on option to their LHDs (cf. B), but without the data transformation component. 
To transform the LiDAR data into the import format of 3DM, the SoS thus uses a customized software component developed by Acme Consultants (cf. C). 
In this scenario, VS2 does not offer the mapping feature as an add-on option. 
Finally, the LiDAR data is imported by 3DM to update its 3D CAD models (cf. D) –- repeatedly synchronized with the maps in Pos (cf. E). 
Depending on the contracts with the suppliers, it is of course possible that Boliden could access the raw data in the databases, e.g., to use the data for other types of analysis.

%Killing some darlings.....
% From a business perspective, Boliden initially buys the mapping feature as an add-on option when purchasing LHD's from VS3 (cf. F) – a feature that implements the whole chain from LiDAR measurements to import into 3DM, but it requires Boliden to pay a subscription fee. Boliden also buys a cheaper but less advanced add-on option from VS1 (cf. H), i.e., delivery of unprocessed LiDAR data. To acquire the necessary data transformation component, Boliden outsources development to ACME Consultants for a fixed cost (cf. area Dev). Both add-on options bought from the vehicle suppliers comes with warranty and support, however limited to the specific mapping feature, i.e., all steps prior to 3DM import. There are no additional running costs from buying the add-on feature.
%
%The vehicle suppliers do not obtain any direct benefits from the SoS itself, but in contrast to Scenario In-house, they are processing LiDAR data – suggesting that the vehicle suppliers employ data analytics, facilitating predictive and preventive maintenance of the LHDs~\cite{levitt_complete_2011}, or at least collecting valuable operations data from the field. In Scenario Add-on, Boliden is no longer sole responsible of the SoS. Instead, the Boliden SoS operations is primarily focused on enabling uninterrupted data flow and transformations from VS1's mapping feature – the VS3 counterpart is an external concern. 

%Regarding the key variation points, we present Scenario Add-on Option as follows:

\subsection{Technical Key Variation Points}

\textbf{IT infrastructure.}
In this scenario, the vehicle suppliers have a critical SoS role as they provide an add-on service to their LHDs. 
The suppliers offering the service would handle the infrastructure up to the point when they deliver their data. 
While VS3 populates 3DM directly, Boliden develops and hosts a customized solution for data provided by VS1. 
Hence, Boliden must operate some IT infrastructure. 

\textbf{Evolution.}
The vehicle suppliers offer a generic feature for the open market. 
Adaptation to specific needs at Boliden are not likely, instead Boliden needs to adapt to the vehicle suppliers. 
The Boliden part of the SoS is rather small and most of its evolution originates in the R\&D of the vehicle suppliers. 
Adding additional Boliden features is difficult, such as novel data analytics. 
On the other hand, LiDAR data from additional vehicle suppliers could be incorporated in the SoS at a later stage, e.g., if VS2 starts offering an analogous mapping feature as an add-on option.

\textbf{ASR.}
Some LiDAR data are stored on VS3's and VS1's LHDs. 
Thus, data security and integrity must be considered from the beginning -- adding security to an already developed system is hard. 
While the LiDAR data stored by the vehicle suppliers are not as sensitive as the final maps, security remains moderately important. 
Efficiency is not an ASR from Boliden's perspective, as this is the concern of the vehicle suppliers who deliver the service with some kind of service level agreement (SLA).
The vehicle suppliers are not used to running a software service that requires monitoring throughout the operation. 
In addition, the vehicle suppliers are not located in the mine, i.e., they are far from the operational environment. 
Consequently, reliability should be considered early and constitutes an ASR in this scenario. 

\subsection{Business Key Variation Points}

\textbf{Inter-company dependency.}
Boliden relies on the vehicle suppliers to offer the mapping feature as an add-on option. 
If this is a feature Boliden prioritizes, they are limited to suppliers actually offering this option.
Furthermore, as there is a customized solution for VS1 LiDAR data, there is a potential lock-in situation. 

\textbf{Upfront investment.}
Boliden pays extra to obtain the mapping feature as an add-on option when purchasing LHDs from VS3 and VS1. 
Furthermore, development of a software component for transformation of LiDAR data from VS1 is outsourced to Acme Consulting.

\textbf{Running costs.}
VS3's more advanced mapping feature offer is provided against a subscription fee. 
On the other hand, Boliden does not pay much to maintain the internal software parts -– although some  resources from Acme Consulting will still be required.

\textbf{Risks.}
Risks in the add-on scenario are shared among Boliden and the vehicle suppliers, both in terms of development costs (upfront investment) and operations costs. 
The risks related to the operation can be regulated in SLAs. 
Even though more data is handled by the suppliers, the different suppliers cannot access each other's data, thus data privacy risks are limited.

\textbf{Innovation platform.}
Boliden is in the same position in terms of driving innovation from a customer perspective~\cite{bosch_speed_2016}. However, in this scenario, there is also a technological innovation drive from the vehicle suppliers.Hence, compared to Scenario In-house, there is a higher chance of innovation at the partners. 

\section{Scenario Additional Service} \label{sec:scen3}% - External Hauling Company Sells Mapping as an Optional Service}
In Scenario Additional Service, Boliden has already outsourced the hauling service to the external company Acme Mining (analogous to a current agreement existing in Kankberg).
Acme Mining has also developed the innovative solution to update maps of underground maps using LiDARs mounted on their LHDs -- and they offer this to Boliden as an additional service. 
Boliden signs a SLA specifying the quality of the mapping service, but do not care which vehicle suppliers' LHDs Acme Mining operate in the mine. 

\begin{figure}
\centering
\includegraphics[width=0.5\textwidth]{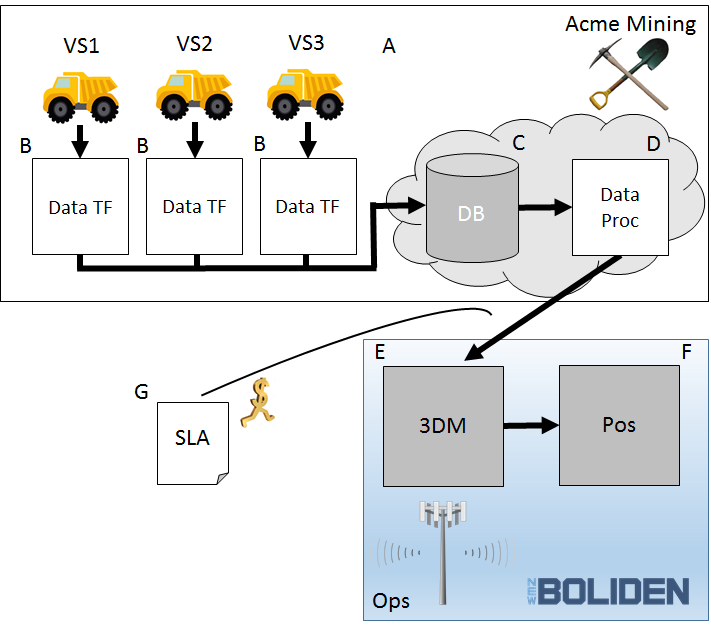}
\caption{Overview of Scenario Additional Service. Acme Mining delivers the hauling service as regulated in an SLA, using LHDs from different vehicle suppliers. Boliden does no development.}
\label{fig:scen3}
\end{figure}

Fig. \ref{fig:scen3} summarizes Scenario Additional Service. 
On top of the figure, Acme Mining operated LHDs from VS1, VS2, and VS3 (cf. A). 
The solution used by Acme Mining contains three different software components for data transformation (cf. B), corresponding to the different LiDAR data formats provided by LHDs from the three vehicle suppliers. 
All LiDAR data are stored by Acme Mining (cf. C), and then they are processed (i.e., filtered and adapted to the 3DM import format, cf D.) prior to transfer to 3DM operated by Boliden. 
Finally, the LiDAR data is imported by 3DM to update its 3D CAD models (cf. E) –- repeatedly synchronized with the maps in Pos (cf. F). 

% Killing more darling / texts
% Looking at the business perspective, Boliden makes only minor upfront investments to realize the SoS benefits. Instead, Acme Mining drives the innovation by providing the mapping as an additional service – not only to Boliden, but to the mining market in general. An SLA (cf. G) between Acme Mining and Boliden regulates frequency and quality of the map updates, and Boliden pays considerable running costs for the service. While the costs are high, Acme Mining takes all risks related to quality of service, and Boliden does not engage in the technical SoS implementation, e.g., whether cloud solutions are used within Acme Mining. In Scenario Additional Service, Boliden does not develop any parts of the SoS themselves. Regarding operations, Boliden can focus on the same constituent systems as today, i.e., 3DM and Pos.

%Regarding the key variation points, we portray Scenario Additional Service as follows:
\subsection{Technical Key Variation Points}

\textbf{IT infrastructure.}
In Scenario Additional Service, the hauling company does not only drive the LHDs, they also operate a map rendering service based on the collected LiDAR data. 
Most of the technical infrastructure will be maintained by the hauling company, who likely uses some a cloud solution. 
Compared to the others, this scenario requires the least from Boliden in terms of IT infrastructure. 

\textbf{Evolution.}
Acme Mining offers mapping as a service to the general market, rather than a solution tailored for Boliden. 
Changes from the vehicle supplier, or modified LiDAR formats, are not likely to be a big problem for Boliden -- it is an issue for the hauling company. 
Hence, from a Boliden perspective, evolution is not much of a concern.

\textbf{ASR.}
Acme Mining is responsible for the development and operations of the solution to render a map from LiDAR data. 
To achieve a scalable solution, they use the same cloud solution for all customers and do not have a separate data center for Boliden, i.e., the cluster for computation and storage is the same across their clients. 
As Acme Mining even renders the map, the final results are more sensitive than just the raw LiDAR data. 
Thus, security in Scenario Additional Service is of utmost importance. 
As Acme Mining is responsible for the operation, efficiency is their responsibility and hence not an ASR from Boliden's perspective. 
However, it is important that efficiency is part of the SLA to ensure an operation suitable for Boliden's needs. 
Reliability follows the same line of reasoning as efficiency, i.e., it is not an ASR but an important concern in the SLA. 

\subsection{Business Key Variation Points}

\textbf{Inter-company dependency.}
The relationship with Acme Mining is fundamental in this scenario –- the SoS exists only as long as they offer the mapping service. 
Switching to a different hauling company risks being a large undertaking, both technically and business-wise. 

\textbf{Upfront investment.}
Boliden needs only minimal initial investments in Scenario Additional Service. 
The required technology development is instead mainly funded by Acme Mining.

\textbf{Running costs.}
Acme Mining charges Boliden for the operation and service offering, i.e., the running costs are comparably high. 
On the other hand, there are minimal maintenance costs for the SoS part operated by Boliden.

\textbf{Risks.}
The investment risks are mainly a concern for Acme Mining. 
Boliden has a single partner with which they contractually regulate the quality of service, transferring risks to the hauling company. 
The biggest risks for Boliden are connected to data privacy, i.e., leaking of sensitive information regarding the ore body and the mining operations.

\textbf{Innovation platform.}
Acme Mining hopes to sell additional services to Boliden (and other customers), thus they work in close collaboration with several vehicle suppliers and have a good opportunity to be innovative. 
However, Acme Mining are not primarily in the mapping business. 
Hence, innovation is largely customer-driven and not technology driven.

\section{Scenario Ecosystem} \label{sec:scen4}% - An Open Software Ecosystem Enables a Third-party Mapping Solution}
The final scenario we describe entails a solution that opens up the business and technical ecosystem.
In Scenario Ecosystem, Boliden has an open technical platform through which services communicate and share data, according to well-defined rules. 
While we present only the realization of the mapping service, the strength of an ecosystem is rather its facilitation of diverse services delivered by different providers, i.e., the possibility for also new solutions to emerge.
There is no such thing as a completely open ecosystem, even though some open source projects come  close. 
In the Boliden case, some parts would be open but not others. 
For example, providing a platform with open APIs and SDK to selected partners is likely, but governance and control would likely not be opened up. 

\begin{figure}
\centering
\includegraphics[width=0.5\textwidth]{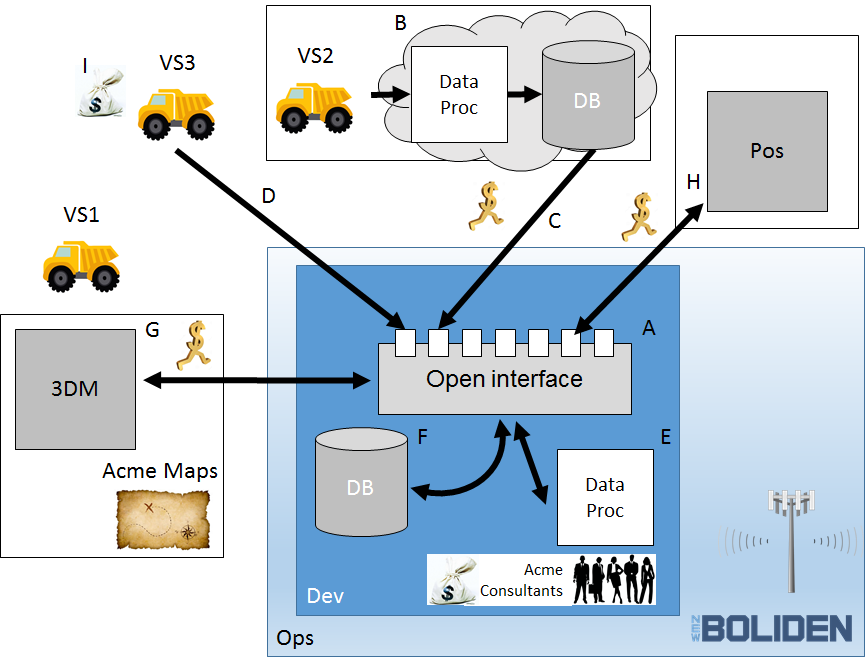}
\caption{Overview of Scenario Ecosystem. Boliden facilitates a semi-open ecosystem that enables a third party mapping solution.}
\label{fig:scen4}
\end{figure}

Fig. \ref{fig:scen4} depicts an overview of Scenario Ecosystem. 
The essential part of the ecosystem is the open interface platform (cf. A), enabling communication within the SoS through the APIs of the constituent systems and software components –- thus simplifying sharing of data and services, as well as integration of new parts. 
VS2 has an advanced LiDAR solution, including data processing and data storage (cf. B). 
The processed LiDAR data from VS2's LHDs are accessible, which they announce using the open interface platform (cf. C). 
Other service providers can then pull LiDAR data from VS2 for various purposes, e.g., map rendering.
Moreover, VS3's LHDs provide access to raw LiDAR data, which other service providers can pull through the open interface (cf. D). 
In practice, Boliden pulls the data to a data processing component (cf. E), and then stores it in a database (cf. F), accessible for other service providers. 
In this Scenario, LHDs from VS1 do not offer any access to LiDAR data. 
3DM recurrently updates its maps by importing LiDAR data through the open interface platform (cf. G), both from Boliden and VS2. 
Finally, Pos synchronizes with the latest spatial data (cf. H) as agreed upon with Boliden. 

% killings some text/darligs... 
% Figure \ref{fig:scen4} also shows key aspects from a business perspective. Boliden outsources development of the open interface platform, i.e., the central component facilitating all communication, to Acme Consultants (cf. Dev). In addition, Acme Consultants develops a data storage solution for Boliden to store LiDAR data from LHDs that do adhere to the 3DM import format. Boliden signs an agreement with VS3 to allow extraction of the LiDAR data, realized through upfront investments (cf. I), in line with Scenario In-house. All other players in the ecosystem are responsible for development of their own software, i.e, Acme Maps, VS2, and Positioning System, but Boliden pays subscription fees for access to services and their corresponding software maintenance.
%
% From an operations point of view, each partner is responsible for their respective parts of the ecosystem. Boliden outsources operations and maintenance of the open interface platform and the internal data storage to Acme Consultants. Compared to the other three scenarios, the emerging SoS is more collaboratively governed, with Boliden oversight though, and the different players are more independent. Although Boliden still has the overall ownership of the SoS, they need less bespoken development. Rather, they would rely on the initiatives and innovation of the different stakeholders and their own capability to provide off the shelf solutions for Boliden. 
 
%Regarding the key variation points, we characterize Scenario Ecosystem as follows:
\subsection{Technical Key Variation Points}

\textbf{IT infrastructure.}
In Scenario Ecosystem, the infrastructure is distributed among the various partners. 
Boliden, with the support from Acme Consultants, operates and maintains the underlying open interface platform. 
The various partners in the ecosystem all operate their own solutions on their own IT infrastructure.

\textbf{Evolution.}
Compared to traditional development, evolution is more dynamic in an ecosystem. 
Each partner continuously develops their solution, and the SoS composition changes over time in a flexible manner. 
Boliden can select the most suitable service from the available solutions in the ecosystem. 

\textbf{ASR.}
In an ecosystem, different systems are interacting in a much more flexible and dynamic manner compared to traditional systems. 
Hence, security is vital to ensure that the right information is shared with the right system.
Furthermore, having an open system where you beforehand do not know exactly which systems will interact with other systems, thus efficiency and reliability must be designed into the common open interface platform. 
If the common platform is not designed properly, then it does not matter how the partaking systems are implemented. 
However, if the platform is designed correctly, then it is up to each participating system/component to implement a sufficient level for the ASRs, of course in accordance with business agreements.  
Consequently, in Scenario Ecosystem, all three quality requirements are ASRs and need to be considered early on. 

\subsection{Business Key Variation Points}

\textbf{Inter-company dependency.} 
Integration of new equipment and services in Scenario Ecosystem should be easy, almost seamless. 
An ecosystem opens up the possibility to easily change service providers. 
Hence, there are few dependencies among the companies.

\textbf{Upfront investment.} 
Boliden makes an upfront investment in the open interface platform supporting the ecosystem, covering both the infrastructure and development outsourced to Acme Consultants. 
Also, development and investing in other pieces of IT infrastructure is needed, leading to moderately high upfront investments.

\textbf{Running costs.} 
Boliden must maintain the infrastructure around the open interface platform. 
The main running costs depend on actual usage of services from ecosystem partners, i.e., VS2, Acme Mining, and Positioning System. 
For example, if Boliden for some reason needs to cut costs, a service can be canceled to save money –- on the other hand, upfront investments might be lost.

\textbf{Risks.}
Investments and operations risks are shared among several partners. 
However, for the emerging properties such as the mapping solution, there is no single partner to agree on a quality of service. 
Consequently, if there is a need to regulate the service in contracts, it requires negotiations with several different companies. 
Data privacy is also a risk, as several partners have access to data and no external partner feels responsible for the integrity of the complete solution –- Boliden must take this role.

\textbf{Innovation platform.} 
An argument for an open ecosystem is that open innovation is supported. 
In essence, the companies collaborate more and drive their core businesses. 
Hence, there will be a good opportunity for innovation, both customer-driven and technology-driven. 

\section{Discussion} \label{sec:disc}
All four scenarios discussed in the previous sections have their individual pros and cons.
Table~\ref{tab:sum} shows a side-by-side comparison of the four scenarios from the perspective of the eight key variation points. 
Based on Table~\ref{tab:sum}, we identify four aspects that are particularly important to consider when realizing a Kankberg SoS.
We believe that \textit{costs}, \textit{control}, \textit{risks}, and \textit{innovation} constitute central elements in trade-offs that must be balanced in future work. 

\begin{table}
\centering
\caption{Comparison of the four system-of-systems scenarios based on the key variation points.}
\includegraphics[width=0.5\textwidth]{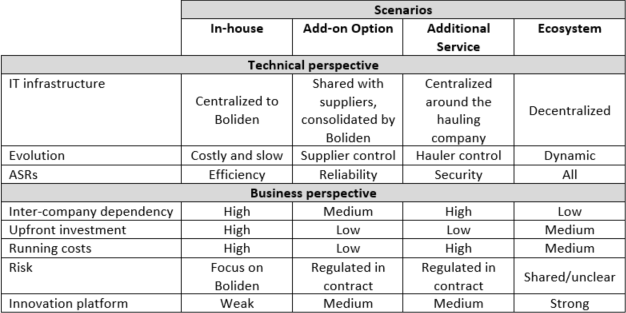}
\label{tab:sum}
\end{table}

\textbf{Control}. Being able to control technical roadmaps, release plans, and quality levels is typically an advantage. 
If a particular SoS aspect is critical, and Boliden has very specific requirements both on content and plan, the control should be prioritized. 
There might also be non-technical aspects for which Boliden has specific requirements, e.g., customer services and 24-hour support to handle any disturbances on the operations. 
That might also require more control from the Boliden side where a more loose cooperation in a software ecosystem might not suffice. 
On the other hand, if Boliden does not have specific requirements, or not the means to invest, then giving up control is not necessarily a disadvantage. 
In Table~\ref{tab:sum}, there tends to be more control of the technical plans on the left-hand side.
To some degree, control means ``do it yourself'', whereas if you want shared costs and/or innovation from outside Boliden, trust in others is essential. 

\textbf{Costs}. Costs are tightly connected to time and quality level, as well as scope of the functionality. 
Discussions on costs encompass both the total costs and upfront costs vs. running costs. 
Sharing costs with suppliers or even competitors means compromising. 
In a software ecosystem, the basic idea is essentially a specialization of actors and a sharing of basic costs~\cite{holmstrom-olsson_strategic_2015}. 
In Table~\ref{tab:sum}, the in-house scenario has the highest upfront cost as well as running costs. 
It is also the scenario in which Boliden has the highest control, hardly anything is shared with others. 
In terms of R\&D cost, the two Scenarios Add-on Option and Additional Service are potentially the cheapest. 
On the other hand, the two scenarios provide less control of what and when the suppliers can deliver. 
A software ecosystem is investment heavy on the short-term, but can be an economical option mid- and long-term.

\textbf{Risks}. The main business risks are related to the supplier network and their ability to deliver products and services. 
In Scenario In-House, Boliden carries all the risks as the overall integrator. 
Hence, both changes in technology and business climate would have to be handled by Boliden. 
Most importantly, there is a large financial risk as the upfront investment cannot be paid back. 
In the other scenarios, a possibility to cancel services remains. 
For Scenario Add-on Feature and Scenario Additional Service, on the other hand, there are risks related to the dependencies to other companies. 
There might be lock-in effects concerning either vehicle suppliers or the hauling company. 
Lock-ins increase the risk of price increases, as switching partners is more difficult when solutions are unique to the partners. 
In Table~\ref{tab:sum}, the two scenarios in the middle are the ones with the lowest risks.
Scenario Ecosystem is unique, as risks are on the one hand shared in the ecosystem, but, on the other hand, there is little or unclear control. 
Hence, risk management becomes a matter of trust among the actors in the software ecosystem. 

\textbf{Innovation}. Boliden is in the mining business and not in the digital innovation business.
However, the digitalization of society will lead to innovation everywhere. 
Even if Boliden is not a technical innovation company, innovation will happen in the company's use cases and business models, i.e., ``customer-driven''~\cite{bosch_speed_2016}. 
Thus, Boliden will also need a foundation for a successful innovation platform. 
In Table~\ref{tab:sum}, the scenarios on the right-hand side enable a more innovation-friendly environment, allowing for technical innovation from partners, as well as from Boliden in terms of customer-driven innovation. 
On the left-hand side, however, the technical innovation from partners is not promoted in those scenarios. 
Especially, the in-house scenario will suffer as the ideas must primarily come from Boliden and not so much from outside influence and new perspectives. 

Thorough analysis and pre-studies are required to identify how to best balance the trade-offs regarding a Kankberg SoS: any decision must be aligned with Boliden's strategic goals. 
It is evident that a customized fully controlled in-house solution will decrease the potential for external innovation, and also be costly –- even more so if risks need to be minimized. 
On the other side of the spectrum, Scenario Ecosystem strongly supports innovation, but the risk that there will be no mature actors in the software ecosystem is high -- and also out of Boliden's control. 
Looking at Scenario Add-on Option and Scenario Additional Feature, these scenarios require less upfront investments, but Boliden is not in control of when such a SoS could be available on the market. 

Finally, we want to highlight that the four scenarios are not mutually exclusive:
a Kankberg SoS could be realized through a combination of in-house solutions, add-on options purchased from vehicle suppliers, and services bought from hauling companies –- such a variety of solutions could even co-exist in a software ecosystem.

\section{Summary and Concluding Remarks} \label{sec:sum}
Ever-increasing production targets under strict safety requirements pushes mining operations toward increased automation. 
In the future, fewer people will work underground -- instead, mining equipment such as LHD trucks will be remote controlled or autonomous. 
As there is no GPS signal in an underground mine, the LHD trucks will navigate using LiDAR. 
The deployment of LHDs equipped with LiDAR sensors will result in large amounts of detailed spatial data, i.e., point clouds from 3D scanning, collected as the LHD trucks drive through drifts and crosscuts.

In the PIMM project, a novel approach to establish reliable wireless communication in the Kankberg mine is explored: an underground 5G network. 
Given a reliable wireless communication network in Kankberg, several new opportunities arise.
Also in the harsh setting of an underground mine, software-intensive solutions relying on high data transfer rates could ripe the benefits of increased digitalization, i.e., embracing the predicted data abundance.

We highlight one possible future approach to utilizing reliable wireless communication: a system-of-systems (SoS) in the Kankberg mine. 
We proposed organizing LHDs equipped with LiDAR and existing map rendering software and indoor positioning systems into a SoS. 
Such a solution could use LiDAR data collected from LHD trucks to continuously update a detailed 3D map of the Kankberg mine. 
The primary benefit would be related to safety of the mining operations, i.e., detecting potential cave-ins, but it would also contribute to productivity as a highly accurate map updated in near real-time would be valuable input to overview mining operations and to plan maintenance activities.

In this paper, we \textit{presented four scenarios illustrating different business models that could enable the Kankberg SoS} (RQ1): 1) an in-house solution, 2) buying mapping capability as an add-on feature from the vehicle suppliers, 3) paying for mapping as an additional service from the hauling service provider, and 4) a software ecosystem. 
For each scenario, we discussed eight key variation points covering both technical and business perspectives. 
We discussed how the different scenarios inevitably leads to trade-off discussions concerning control, costs, risks, and innovation. 
%An in-house solution would provide full control of both features and risks, but be costly and provide a weaker innovation platform.
It is evident that the four scenarios all have their pros and cons, and the SoS requirements are affected accordingly (RQ2). 
We argue that from Boliden's perspective, \textit{the architecturally significant requirements (ASRs)}~\cite{chen_characterizing_2013} for the Scenarios In-house, Add-on Option, and Additional Service would be efficiency, reliability, and security, respectively. 
On the other hand, \textit{if Boliden would be the platform provider in Scenario Ecosystem, the company must carefully develop all these quality requirements}.

We propose three ways forward to realized a SoS in Kankberg, all involving a closer collaboration with vehicle suppliers. 
First, a LiDAR mapping feasibility study is needed to develop a proof-of-concept prototype. 
Boliden could mount existing LiDAR sensors on manually operated LHDs, and develop a data extraction system that also feeds the mapping solutions.
%It should include not only the technical challenges but also the potential partners to work through a full-fledged business scenario. 
Second, a software ecosystem pre-study should be conducted to investigate both the requirements for a robust ecosystem platform, as well as the potential to establish a group of actors, i.e., a community. 
Is it realistic for the future mining industry to use a software ecosystem paradigm in the IT infrastructure and innovation platform? 
Third, a more thorough study on the SoS is required. 
In a SoS, operation and control are different compared to traditional systems. 
It is not obvious how to set up the IT infrastructure and enterprise architecture to ensure efficiency, quality, and maintainability. 
An in-depth study on the current IT infrastructure, combined with elicitation of infrastructure requirements needed to enable a SoS, could provide necessary answers on what needs to be changed in Kankberg. 

Analogous to other traditional industries, the future of mining will be increasingly dependent on software. 
All future projections point toward digitalization, shared data, and increased automation. 
This new environment will fundamentally change business models, and software will be the main driver in R\&D investments. 
Given reliable wireless communication, software will be the glue that enables SoS in mines such as Kankberg –- the burning question is: which actor will act first to champion their preferred  business model? 

\section*{Acknowledgment}
This work was funded by VINNOVA, the Swedish Agency for Innovation Systems within the PIMM project, Pilot for Industrial Mobile communication in Mining.



%\bibliographystyle{IEEEtran}

\bibliographystyle{IEEEtran}
\bibliography{re2017}

\end{document}